# Determination of energy flux rate in homogeneous ferrohydrodynamic turbulence using two-point statistics

Sukhdev Mouraya[*] and Supratik Banerjee[†]

*Department of Physics,*

*Indian Institute of Technology Kanpur, 208016, India.*

(Dated: July 1, 2019)

i

## Abstract

In ferrofluids the suspended ferromagnetic particles agglomerate due to the interaction between the particle magnetic moment and the external magnetic field, which in turn, influences the turbulence and relaxation time. The relaxation time becomes large when we are considering turbulence drag force instead of viscous drag force in Brownian motion. We investigate that the total energy conservation in ferrofluids taking into interaction between the external magnetic field and the ferrofluid. Using two point statistics we can formulate exact relations in the inertial zone of the turbulence. This exact relation shows that $(\mathbf{u} \times \boldsymbol{\omega})$, $(\mathbf{M} \times \mathbf{H})$, $\mathbf{M} \cdot \nabla \mathbf{H}$ and $(\boldsymbol{\omega} \times \mathbf{M})$ play the major role in energy cascading.

## I. INTRODUCTION

A ferrofluid is a suspension of solid magnetic particles (magnetite) in dielectric liquids (e.g. water, oil, organic solvent), often called the carrier fluid. The diameter of the suspended particles are of the order of 10 $nm$ and hence each particle contains roughly one single magnetic domain [1]. These magnetic particles are free to move and rotate in the carrier medium. Due to their very small size, the particles interact with each other when subject to a magnetic field. However, due to the Brownian motion, they do not settle under the influence of external magnetic field. The ferrofluid particles are usually coated with a surfactant, which are long chained molecules having a polar head and a non-polar tail or vice versa (e.g. oleic acid, tetramethylammonium hydroxide etc.). This coating prevents the particles from (i) immediate agglomeration among themselves by producing steric hinderance and (ii) being extracted from the fluid, when the external field is sufficiently strong. This surfactant partially overcomes the attractive Van der Waals and magnetic forces between the particles with electrostatic repulsion [2]. When an external magnetic field is applied, the magnetic moments of these ferromagnetic particles, which are assumed to be fixed with each particle, immediately try to be oriented along the field direction. When the field is removed, these magnetic moments quickly randomize leading to a zero net magnetization [3, 4]. This property, along with the fact that ferrofluids possess magnetic susceptibilities ($\sim 0.5$) almost three orders of magnitude larger than the common paramagnetic salt, classify the ferrofluids as superparamagnetic fluids [1]. Owing to their high controllability by external magnetic fields and convective effects in microgravity environments, ferrofluids find a broad range of applications starting from hermetic seals in pumps, rotary seals of computer hard-drives to the loudspeakers to extract heat from voice coil [1, 5, 6]. Ferrofluids are also considered as effective carrier of concentrated medications to specific location in the body [7].

---

$^{*}$ sukhdev@iitk.ac.in

$^{\dagger}$ sbanerjee@iitk.ac.in

In this article, we mainly investigate the comportment of a ferrofluid flow in the external magnetic field ($\mathbf{H}$) gradient when $\mathbf{H}$ is steady *i.e.* time independent (see figure). Upon the application of steady $\mathbf{H}$, the particle magnetic moments start getting oriented along it thereby giving an equilibrium state in a characteristic time $\tau$, called the relaxation time [2]. In the equilibrium state all the particles get settled and oriented along $\mathbf{H}$ (see figure). In order to achieve a sustaining ferrofluid flow, the process of agglomeration of the particles needs to be slow. As we discuss in Sec II, this can be achieved through the generation of turbulence in a ferrofluid flow. Turbulence modifies the nature of the viscous drag and increases the relaxation time of particles to settle. In addition, the interaction between the ferrofluid and $\mathbf{H}$ considerably influences the properties of turbulence. Although a large number of studies [8–11] have been accomplished for the laminar Poiseuille flow of ferrofluids and more particularly the drag reduction in the presence of rotating and oscillating magnetic fields, only a few works have been done to study the turbulence in such fluids. Some of them are dedicated to the study the effects of pressure drop in the pipe flow of ferrofluids [12, 13] whereas some investigate systematically the onset of turbulence in Taylor-Couette ferrofluid flow which takes place at Reynolds numbers at least ten times lower than that of the other fluids [14]. The study of completely developed homogeneous turbulence in ferrofludic flows was carried out, for the first time, by Schumacher *et al.* [1]. Starting from the very basic equations of ferrofluids, using one-point Reynolds decomposition method, the authors derived the expression for average dissipation rate of the translational kinetic energy of the fluid, the rotational kinetic energy of the suspended ferrofluid particles and the internal energy. They also identified the terms which represent the conversion of one type of energy into the other. The analytical part were then numerically calculated using DNS. The external field $\mathbf{H}$ was assumed to be steady throughout. The effect of a spatially uniform but oscillating $\mathbf{H}$ on homogeneous ferrofluid turbulence was studied later by the same authors [15] and an increased rate of energy loss was reported. The study also showed a direct dependence of turbulent properties on the oscillation frequency of the external magnetic field and the choice of magnetization equation. The properties of ferrohydrodynamic turbulence in a channel flow under both steady and oscillating magnetic fields were also studied numerically using DNS and the results were found to match satisfactorily with the $k - \epsilon$ model of turbulence adapted to ferrofluids [16]. Despite these extensive studies, to our knowledge, no exact relation has been derived for homogeneous ferrofluid turbulence using two-point statistics. Exact relations are crucial as they can directly express the average energy dissipation rate in terms of two-point fluctuations which corresponds to the quantities as a function of the length scale. The first exact relation for incompressible hydrodynamic turbulence was derived by [? ]. Later such exact relations were derived for incompressible magnetohydrodynamics [? ? ] and also for various types of compressible turbulence [? ? ]. Very recently [? ] and Banerjee and Galtier [17] have proposed an alternative method of derivation of the exact relations. This method is found to have several advantages and the final exact relation becomes more compact than the previous ones. This method was successfully generalized for compressible turbulence of both neutral and plasma fluids [18].

Following [17, 18], in this paper, we derive an exact relation corresponding to the total energy conservation in the so-called inertial zone of incompressible ferrofluid turbulence. The paper is organised as follows. The governing equations of dynamics are presented in Sec. II. In Sec. III , the total energy consevation is shown and the following section contains the detailed derivation of an exact relation using two-point statistics. Finally in Sec. V, we summerize the results and conclude.

## II. BASIC EQUATIONS OF DYNAMICS

The basic equations of dynamics for ferrofluids consist of the linear momentum equation of the fluid, the internal angular momentum equation of the suspended ferromagnetic particles and the magnetization equation. The governing equation for linear momentum evolution is derived from the theory of structured continua [1, 3] and can be written as in the following:

$$\rho \left[ \frac{\partial \mathbf{u}}{\partial t} + (\mathbf{u} \cdot \nabla)\mathbf{u} \right] = -\nabla p + \mu \nabla^2 \mathbf{u} + \mu_0(\mathbf{M} \cdot \nabla \mathbf{H}) - \zeta \nabla \times ((\nabla \times \mathbf{u}) - 2\boldsymbol{\omega}) + \mathbf{f}_u, \qquad (1)$$

where $\rho$ is the density of ferrofluid, $\mathbf{u}$ the average velocity, $p$ the dynamic pressure, $\mu$ the dynamic viscosity, $\zeta$ the vortex viscosity, $\mu_0(\mathbf{M} \cdot \boldsymbol{\nabla}\mathbf{H})$ the magnetic body force, $\mu_0$ the free space permeability, $\mathbf{M}$ the magnetization vector, $\mathbf{H}$ the external magnetic field, $\boldsymbol{\omega}$ the local ferrofluid rotation rate and $\mathbf{f}_u$ the stationary large scale force. The evolution equation of the internal angular momentum, which is the total angular momentum minus the moment of the total linear momentum [3], is given by,

$$\rho I \left[ \frac{\partial \boldsymbol{\omega}}{\partial t} + (\mathbf{u} \cdot \nabla) \boldsymbol{\omega} \right] = \eta \nabla^2 \boldsymbol{\omega} + \mu_0 (\mathbf{M} \times \mathbf{H}) + 2\zeta((\nabla \times \mathbf{u}) - 2\boldsymbol{\omega}), \qquad (2)$$

where $I$ is the moment of inertia of ferrofluid particle, $\eta$ the spin viscosity and $\mu_0(\mathbf{M} \times \mathbf{H})$ represents the magnetic body couple force. In addition to these two equations, we assume incompressibility which gives $\nabla \cdot \mathbf{u} = 0$. Again the Maxwell's equations for ferrofluid with no current are written as

$$\nabla \times \mathbf{H} = 0; \qquad \nabla \cdot \mathbf{B} = 0, \qquad (3)$$

and the relation between $\mathbf{M}$, $\mathbf{H}$ and $\mathbf{B}$ is

$$\mathbf{B} = \mu_0 (\mathbf{H} + \mathbf{M}). \qquad (4)$$

For the magnetization equation, we have more than one choice. Whereas the original one was proposed by Shliomis [4], two other magnetization equations were proposed later by **?** ] and **?** ] using Fokker Planck equation and irreversible thermodynamics respectively. Here as a first step, we use the equation proposed by Shliomis [4]:

$$\frac{\partial \mathbf{M}}{\partial t} + (\mathbf{u} \cdot \nabla) \mathbf{M} = \boldsymbol{\omega} \times \mathbf{M} - \frac{1}{\tau} (\mathbf{M} - \mathbf{M}_0), \qquad (5)$$

where $\tau$ is the relaxation time, $\boldsymbol{\omega} \times \mathbf{M}$ represents the rate of change in magnetization due to rotation of magnetic particles, $\frac{1}{\tau}(\mathbf{M} - \mathbf{M}_0)$ represents the change in magnetization towards an equilibrium magnetization ($\mathbf{M}_0$) via relaxation. This equilibrium magnetization is given by:

$$\mathbf{M}_0 = M_s L(\xi) \frac{\mathbf{H}}{H}, \qquad (6)$$

where

$$L(\xi) = \coth(\xi) - \frac{1}{\xi} \quad \text{and} \quad \xi = \frac{\mu_0 m H}{k_B T}, \qquad (7)$$

and $M_s$, m are the magnitudes of saturation magnetization and magnetic moment of a single particle.

Here we are discussing ferrofluidic flow in turbulent regime. Therefore instead of taking viscous drag force we use the turbulent drag force. The viscous drag force [19] is:

$$F^v = 6\pi \mu r u, \qquad (8)$$

where $r$ is the radius of the particle. For this Brownian relaxation time [1] is:

$$\tau_B^v = 3\frac{V_p\mu}{k_BT}, \tag{9}$$

where $V_p$ is the hydrodynamic volume of the magnetic particle, $k_BT$ is the thermal energy. Now the turbulent drag force [19, 20] is:

$$F^t = \frac{1}{2}\rho u^2 C_d A, \tag{10}$$

where $C_d$ is the drag coefficient which is more or less constant at high Reynolds number and $A$ is the cross sectional area. So the diffusion coefficient($D$) is given by:

$$D = \frac{2k_BT}{\rho u C_d A}. \tag{11}$$

Hence the Einstein equation for relaxation time of Brownian motion is modified as

$$\tau_B^t = \frac{\rho \Delta^2 u C_d A}{4k_BT}. \tag{12}$$

Here relaxation time $(\tau \approx \tau_B^t)$ is directly propotional to the velocity which is high for turbulent region so $\frac{1}{\tau}$ is very small. Hence the magnetization equation 5 can be written as:

$$\frac{\partial \mathbf{M}}{\partial t} + (\mathbf{u} \cdot \nabla)\mathbf{M} = \boldsymbol{\omega} \times \mathbf{M}. \tag{13}$$

## III. CONSERVATION OF ENERGY IN TURBULENT RESIGME

The evolution equation for translational kinetic energy is given by the use of equations 1, 2 and **??**, which is given below:

$$\frac{\partial}{\partial t}\left(\frac{u^2}{2}\right) + \nabla \cdot \left(\frac{u^2}{2}\mathbf{u}\right) = -\nabla \cdot (p\mathbf{u}) + \nu\mathbf{u} \cdot \nabla^2\mathbf{u} + 2\zeta\mathbf{u} \cdot (\nabla \times \boldsymbol{\omega}) - \zeta\mathbf{u} \cdot (\nabla \times (\nabla \times \mathbf{u}))$$
$$+ \mu_0\mathbf{u} \cdot (\mathbf{M} \cdot \nabla\mathbf{H}) + \mathbf{u} \cdot \mathbf{f}_u, \tag{14}$$

and the evolution equation for rotational kinetic energy can be written as:-

$$I\frac{\partial}{\partial t}\left(\frac{\omega^2}{2}\right) + \nabla \cdot (I\frac{\omega^2}{2}\mathbf{u}) = \eta\boldsymbol{\omega} \cdot \nabla^2\boldsymbol{\omega} + 2\zeta\boldsymbol{\omega} \cdot (\nabla \times \mathbf{u} - 2\boldsymbol{\omega}) + \mu_0\boldsymbol{\omega} \cdot (\mathbf{M} \times \mathbf{H}), \tag{15}$$

where we have used the relation $\boldsymbol{\omega}(\mathbf{u} \cdot \nabla)\boldsymbol{\omega} = \mathbf{u} \cdot \nabla\left(\omega^2/2\right)$.

Integrating over entire volume of equation 14 and 15, and on adding we get,

$$\frac{\partial}{\partial t}\left(\int \frac{u^2}{2} + I\frac{\boldsymbol{\omega}^2}{2}\right) d\tau = -\int \nabla \cdot \left(\frac{u^2}{2} + I\frac{\boldsymbol{\omega}^2}{2} + p\right)\mathbf{u}d\tau + \mu_0 \int \left[\mathbf{u} \cdot (\mathbf{M} \cdot \nabla \mathbf{H}) + \cdot(\mathbf{M} \times \mathbf{H})\right] d\tau$$

$$+ \int \left[\nu\mathbf{u} \cdot \nabla^2\mathbf{u} + \eta\boldsymbol{\omega} \cdot \nabla^2\boldsymbol{\omega}\right] d\tau - \zeta \int \left[\mathbf{u} \cdot \nabla \times (\nabla \times \mathbf{u} - 2\boldsymbol{\omega}) - \boldsymbol{\omega} \cdot (\nabla \times \mathbf{u} - 2\boldsymbol{\omega})\right] d\tau \quad (16)$$

For invisid fluid $\mu = 0$. Schumacher et al. (2003) [13] estimated that $\zeta/\mu = 0.50$ and $\eta = 2 \times 10^{-15}$ kg m $s^{-1}$, so one can set $\zeta = 0$ and $\eta = 0$. By using Gauss divergence equation 16, we get:

$$\frac{\partial}{\partial t}\left(\int \frac{u^2}{2} + I\frac{\boldsymbol{\omega}^2}{2}\right) d\tau = \mu_0 \int \mathbf{u} \cdot (\mathbf{M} \cdot \nabla)\mathbf{H}d\tau + \mu_0 \int \boldsymbol{\omega} \cdot (\mathbf{M} \times \mathbf{H})d\tau \quad (17)$$

Since the magnetic field is time independent, so $\partial\mathbf{H}/\partial t = 0$. Rate of work done on the system by external magnetic field is :-

$$\frac{d}{dt}\int W d\tau = \frac{\partial}{\partial t}\int W d\tau = \int \mathbf{H} \cdot \frac{\partial\mathbf{B}}{\partial t}d\tau = \mu_0 \int \mathbf{H} \cdot \frac{\partial\mathbf{M}}{\partial t}d\tau = \frac{\partial}{\partial t}\int \mu_0(\mathbf{H} \cdot \mathbf{M})d\tau$$

$$= \mu_0 \int \mathbf{H} \cdot (-(\mathbf{u} \cdot \nabla\mathbf{M}) + \boldsymbol{\omega} \times \mathbf{M})d\tau$$

$$= -\mu_0 \int \nabla \cdot (\mathbf{M}(\mathbf{u} \cdot \mathbf{H}))d\tau + \mu_0 \int \mathbf{u} \cdot (\mathbf{M} \cdot \nabla)\mathbf{H}d\tau + \mu_0 \int \nabla \cdot ((\mathbf{u} \times \mathbf{M}) \times \mathbf{H})d\tau$$

$$+ \mu_0 \int \mathbf{H} \cdot (\omega \times \mathbf{M})d\tau$$

$$= \mu_0 \int \mathbf{u} \cdot (\mathbf{M} \cdot \nabla)\mathbf{H}d\tau + \mu_0 \int \mathbf{H} \cdot (\omega \times \mathbf{M})d\tau \quad (18)$$

Here relation $\mathbf{H} \cdot (\mathbf{u} \cdot \nabla)\mathbf{M} = \nabla \cdot (\mathbf{M}(\mathbf{u} \cdot \mathbf{H})) - \mathbf{u} \cdot (\mathbf{M} \cdot \nabla)\mathbf{H} - \nabla \cdot ((\mathbf{u} \times \mathbf{M}) \times \mathbf{H})$ has been used. Since here neither energy injected nor energy dissipation, hence the work done by magnetic field is consumed in two parts translational and rotational kinetic energy so:

$$\frac{\partial}{\partial t}\left(\int \frac{u^2}{2} + I\frac{\omega^2}{2}\right) d\tau - \frac{\partial}{\partial t}\int W d\tau = \frac{\partial}{\partial t}\left(\int \frac{u^2}{2} + I\frac{\omega^2}{2} - \mu_0\mathbf{H} \cdot \mathbf{M}\right) d\tau = 0 \quad (19)$$

Here we can't say that total energy is constant because the work done by the applied magnetic field is $\mathbf{H} \cdot d\mathbf{B}$ not $\mu_0(\mathbf{H} \cdot \mathbf{M})$. This comes because magnetic field($\mathbf{H}$) is time independent. Hence one can say that the rate of total energy is equals the above value. So one can write:

$$\int \left(\frac{u^2}{2} + I\frac{\omega^2}{2} - \mu_0(\mathbf{H} \cdot \mathbf{M})\right) d\tau = Constant \quad (20)$$

## IV. DERIVATION OF EXACT RELATION

In conservation of energy we see that one can write total energy is conserve. Hence energy flux rate of which can be used in correlation function. The two-point symmetric correlation function of energy [17] can be written as:

$$\mathcal{R} = \frac{R_E + R'_E}{2} = \frac{1}{2} \langle (\mathbf{u} \cdot \mathbf{u}' + I\boldsymbol{\omega} \cdot \boldsymbol{\omega}' - \mu_0(\mathbf{H} \cdot \mathbf{M}' + \mathbf{H}' \cdot \mathbf{M})\rangle \tag{21}$$

For statistical stationary state-

$$\frac{\partial(R_E + R'_E)}{\partial t} = 0 \tag{22}$$

The evolution equation of the correlation function by using the basic equations of dynamics [17] is:

$$\begin{aligned}
\frac{\partial \mathcal{R}}{\partial t} &= \frac{1}{2}\frac{\partial}{\partial t}\langle(\mathbf{u} \cdot \mathbf{u}' + I\boldsymbol{\omega} \cdot \boldsymbol{\omega}' - \mu_0(\mathbf{H} \cdot \mathbf{M}' + \mathbf{H}' \cdot \mathbf{M})\rangle \\
&= \frac{1}{2}\langle(\mathbf{u} \times \boldsymbol{\Omega}) \cdot \mathbf{u}' + (\mathbf{u}' \times \boldsymbol{\Omega}') \cdot \mathbf{u} - I(\mathbf{u} \cdot \nabla)\boldsymbol{\omega} \cdot \boldsymbol{\omega}' - I(\mathbf{u}' \cdot \nabla')\boldsymbol{\omega}' \cdot \boldsymbol{\omega}\rangle \\
&+ \frac{1}{2}\mu_0\langle(\mathbf{M} \times \mathbf{H}) \cdot \boldsymbol{\omega}' + (\mathbf{M}' \times \mathbf{H}') \cdot \boldsymbol{\omega} - \mathbf{H}' \cdot (\boldsymbol{\omega} \times \mathbf{M}) - \mathbf{H} \cdot (\boldsymbol{\omega}' \times \mathbf{M}') + \mathbf{u}' \cdot (\mathbf{M} \cdot \nabla)\mathbf{H}\rangle \\
&+ \frac{1}{2}\mu_0\langle\mathbf{u} \cdot (\mathbf{M}' \cdot \nabla')\mathbf{H}' + \mathbf{H}' \cdot (\mathbf{u} \cdot \nabla)\mathbf{M} + \mathbf{H} \cdot (\mathbf{u}' \cdot \nabla')\mathbf{M}'\rangle + D_u + D_\zeta + D_w + F_u
\end{aligned} \tag{23}$$

where $D_u$, $D_w$ represents kinematic viscous dissipation, $D_\zeta$ represents vortex dissipation and $F_u$ represents the forcing contribution, which are given below:

$$D_u = \nu\langle\mathbf{u} \cdot \nabla'^2\mathbf{u}' + \mathbf{u}' \cdot \nabla^2\mathbf{u}\rangle, \tag{24}$$

$$D_\zeta = -\zeta\langle\mathbf{u}' \cdot \nabla \times (\boldsymbol{\Omega} - 2\boldsymbol{\omega}) + \mathbf{u} \cdot \nabla' \times (\boldsymbol{\Omega}' - 2\boldsymbol{\omega}') - 2(\boldsymbol{\omega}' \cdot (\boldsymbol{\Omega} - 2\boldsymbol{\omega}) + \boldsymbol{\omega} \cdot (\boldsymbol{\Omega}' - 2\boldsymbol{\omega}'))\rangle, \tag{25}$$

$$D_w = \eta\langle\boldsymbol{\omega}' \cdot \nabla^2\boldsymbol{\omega} + \boldsymbol{\omega} \cdot \nabla'^2\boldsymbol{\omega}'\rangle, \tag{26}$$

$$F_u = \langle\mathbf{u} \cdot \mathbf{f}'_u + \mathbf{u}' \cdot \mathbf{f}_u\rangle \tag{27}$$

For incompressible and statistical homogeneous system we can show the relations $\left\langle\nabla \cdot \left(\frac{u^2}{2}\mathbf{u}'\right)\right\rangle = \left\langle\nabla' \cdot \left(\frac{u^2}{2}\mathbf{u}'\right)\right\rangle = 0$ and $\langle\mathbf{u}' \cdot \nabla p\rangle = \langle\nabla \cdot (p\mathbf{u}')\rangle = \langle\nabla' \cdot (p\mathbf{u}')\rangle = 0$.

Using statistical homogeneity one can prove that

$$\langle(\mathbf{u} \times \boldsymbol{\Omega}) \cdot \mathbf{u}' + (\mathbf{u}' \times \boldsymbol{\Omega}') \cdot \mathbf{u}\rangle = -\langle\delta(\mathbf{u} \times \boldsymbol{\Omega}) \cdot \delta\mathbf{u}\rangle, \tag{28}$$

$$\langle\mathbf{M} \times \mathbf{H}) \cdot \boldsymbol{\omega}' + (\mathbf{M}' \times \mathbf{H}') \cdot \boldsymbol{\omega} - \mathbf{H}' \cdot (\boldsymbol{\omega} \times \mathbf{M}) - \mathbf{H} \cdot (\boldsymbol{\omega}' \times \mathbf{M}')\rangle = \langle\delta\mathbf{H} \cdot \delta(\boldsymbol{\omega} \times \mathbf{M})\rangle$$
$$- \langle\delta\boldsymbol{\omega} \cdot \delta(\mathbf{M} \times \mathbf{H})\rangle, \tag{29}$$

$$\langle-(\mathbf{u} \cdot \boldsymbol{\nabla})\boldsymbol{\omega} \cdot \boldsymbol{\omega}' - (\mathbf{u}' \cdot \boldsymbol{\nabla}')\boldsymbol{\omega}' \cdot \boldsymbol{\omega}\rangle = \langle\delta\boldsymbol{\omega} \cdot \delta(\mathbf{u} \cdot \boldsymbol{\nabla}\boldsymbol{\omega})\rangle, \tag{30}$$

here relation $\langle\boldsymbol{\omega}\cdot(\mathbf{u}\cdot\nabla\boldsymbol{\omega})\rangle=\left\langle\mathbf{u}\cdot\nabla\left(\frac{\omega^2}{2}\right)\right\rangle=\left\langle\nabla\cdot\left(\frac{\omega^2}{2}\mathbf{u}\right)\right\rangle=\left\langle\nabla'\cdot\left(\frac{\omega^2}{2}\mathbf{u}\right)\right\rangle=0$ has been
used and the remaining term can be provided by using the relations:

$$\langle\delta\mathbf{u}\cdot\delta((\mathbf{M}\cdot\nabla)\mathbf{H})\rangle=\langle(\mathbf{u}'-\mathbf{u})\cdot((\mathbf{M}'\cdot\nabla')\mathbf{H}'-(\mathbf{M}\cdot\nabla)\mathbf{H})\rangle$$
$$=\langle\mathbf{u}'\cdot(\mathbf{M}'\cdot\nabla')\mathbf{H}'-\mathbf{u}\cdot(\mathbf{M}'\cdot\nabla')\mathbf{H}'-\mathbf{u}'\cdot(\mathbf{M}\cdot\nabla)\mathbf{H}+\mathbf{u}\cdot(\mathbf{M}\cdot\nabla)\mathbf{H}\rangle,\qquad(31)$$

$$\langle\delta\mathbf{H}\cdot\delta((\mathbf{u}\cdot\nabla)\mathbf{M})\rangle=\langle(\mathbf{H}'-\mathbf{H})\cdot((\mathbf{u}'\cdot\nabla')\mathbf{M}'-(\mathbf{u}\cdot\nabla)\mathbf{M})\rangle$$
$$=\langle\mathbf{H}'\cdot(\mathbf{u}'\cdot\nabla')\mathbf{M}'-\mathbf{H}\cdot(\mathbf{u}'\cdot\nabla')\mathbf{M}'-\mathbf{H}'\cdot(\mathbf{u}\cdot\nabla)\mathbf{M}+\mathbf{H}\cdot(\mathbf{u}\cdot\nabla)\mathbf{M}\rangle$$
$$=\langle\nabla'\cdot((\mathbf{M}'\cdot\mathbf{H}')\mathbf{u}')-\mathbf{M}'\cdot(\mathbf{u}'\cdot\nabla')\mathbf{H}'-\mathbf{H}\cdot(\mathbf{u}'\cdot\nabla')\mathbf{M}'-\mathbf{H}'\cdot(\mathbf{u}\cdot\nabla)\mathbf{M}\rangle$$
$$+\langle\nabla\cdot((\mathbf{M}\cdot\mathbf{H})\mathbf{u})-\mathbf{M}\cdot(\mathbf{u}\cdot\nabla)\mathbf{H}\rangle,\qquad(32)$$

and some relations which we will use further:

$$\nabla\cdot((\mathbf{M}\cdot\mathbf{H})\mathbf{u})=\mathbf{M}\cdot(\mathbf{u}\cdot\nabla)\mathbf{H}+\mathbf{H}\cdot(\mathbf{u}\cdot\nabla)\mathbf{M},\qquad(33)$$

$$\langle\nabla\cdot((\mathbf{M}\cdot\mathbf{H})\mathbf{u})\rangle=-\nabla_l\cdot\langle(\mathbf{M}\cdot\mathbf{H})\mathbf{u}\rangle=\langle\nabla'\cdot((\mathbf{M}\cdot\mathbf{H})\mathbf{u})\rangle=0,\qquad(34)$$

$$\mathbf{u}\cdot(\mathbf{M}\cdot\nabla)\mathbf{H}-\mathbf{M}\cdot(\mathbf{u}\cdot\nabla)\mathbf{H}=(\mathbf{M}\times\mathbf{u})\cdot(\nabla\times\mathbf{H})=0.\qquad(35)$$

Adding equation 31 and 32 and using the above relations one can get:

$$\langle\delta\mathbf{u}\cdot\delta((\mathbf{M}\cdot\nabla)\mathbf{H})+\delta\mathbf{H}\cdot\delta((\mathbf{u}\cdot\nabla)\mathbf{M})\rangle=\langle-\mathbf{u}\cdot(\mathbf{M}'\cdot\nabla')\mathbf{H}'-\mathbf{u}'\cdot(\mathbf{M}\cdot\nabla)\mathbf{H}\rangle$$
$$+\langle-\mathbf{H}\cdot(\mathbf{u}'\cdot\nabla')\mathbf{M}'-\mathbf{H}'\cdot(\mathbf{u}\cdot\nabla)\mathbf{M}+\mathbf{u}'\cdot(\mathbf{M}'\cdot\nabla')\mathbf{H}'-\mathbf{M}'\cdot(\mathbf{u}'\cdot\nabla')\mathbf{H}'+\mathbf{u}\cdot(\mathbf{M}\cdot\nabla)\mathbf{H}\rangle$$
$$+\langle-\mathbf{M}\cdot(\mathbf{u}\cdot\nabla)\mathbf{H}+\nabla'\cdot((\mathbf{M}'\cdot\mathbf{H}')\mathbf{u}')+\nabla\cdot((\mathbf{M}\cdot\mathbf{H})\mathbf{u})\rangle$$
$$=\langle-\mathbf{u}\cdot(\mathbf{M}'\cdot\nabla')\mathbf{H}'-\mathbf{u}'\cdot(\mathbf{M}\cdot\nabla)\mathbf{H}-\mathbf{H}\cdot(\mathbf{u}'\cdot\nabla')\mathbf{M}'-\mathbf{H}'\cdot(\mathbf{u}\cdot\nabla)\mathbf{M}+\nabla\cdot((\mathbf{M}\cdot\mathbf{H})\mathbf{u})\rangle$$
$$+\langle\nabla'\cdot((\mathbf{M}'\cdot\mathbf{H}')\mathbf{u}')+(\mathbf{M}\times\mathbf{u})\cdot(\nabla\times\mathbf{H})+(\mathbf{M}'\times\mathbf{u}')\cdot(\nabla'\times\mathbf{H}')\rangle$$
$$=\langle-\mathbf{u}\cdot(\mathbf{M}'\cdot\nabla')\mathbf{H}'-\mathbf{u}'\cdot(\mathbf{M}\cdot\nabla)\mathbf{H}-\mathbf{H}\cdot(\mathbf{u}'\cdot\nabla')\mathbf{M}'-\mathbf{H}'\cdot(\mathbf{u}\cdot\nabla)\mathbf{M}\rangle.\qquad(36)$$

Combining the equations 28, 29, 36, 38 and using in equation 23, we get:

$$\frac{\partial\mathcal{R}}{\partial t}=\frac{1}{2}\langle-\delta(\mathbf{u}\times\boldsymbol{\Omega})\cdot\delta\mathbf{u}+I\delta\boldsymbol{\omega}\cdot\delta(\mathbf{u}\cdot\nabla\boldsymbol{\omega})+\mu_0(\delta\mathbf{H}\cdot\delta(\boldsymbol{\omega}\times\mathbf{M})-\delta\boldsymbol{\omega}\cdot\delta(\mathbf{M}\times\mathbf{H}))\rangle$$
$$-\frac{1}{2}\mu_0\langle\delta\mathbf{u}\cdot\delta((\mathbf{M}\cdot\nabla)\mathbf{H})+\delta\mathbf{H}\cdot\delta((\mathbf{u}\cdot\nabla)\mathbf{M})\rangle+D_w++D_\varsigma+D_u+F_u\qquad(37)$$

For statistical stationary state $\partial_t(R_E+R'_E)=0$ and for inertial range we can neglect the
dissipation term then we derive the following exact relation:

$$2\varepsilon=\langle\delta(\mathbf{u}\times\boldsymbol{\Omega})\cdot\delta\mathbf{u}-I\delta\boldsymbol{\omega}\cdot\delta(\mathbf{u}\cdot\nabla\boldsymbol{\omega})+\mu_0(\delta\mathbf{u}\cdot\delta((\mathbf{M}\cdot\nabla)\mathbf{H})+\delta\mathbf{H}\cdot\delta((\mathbf{u}\cdot\nabla)\mathbf{M}))$$
$$+\langle-\delta\mathbf{H}\cdot\delta(\boldsymbol{\omega}\times\mathbf{M})+\delta\boldsymbol{\omega}\cdot\delta(\mathbf{M}\times\mathbf{H})\rangle,\qquad(38)$$

where $\varepsilon$ is mean energy flux rate, given by $F_u = \varepsilon$, large scale forcing. Eqn (38) is the central result of this paper. This expresses the average flux rate of total energy ($\varepsilon$) for completely developed turbulence in an incompressible ferrofluid in terms of two-point increments of different fluid and electromagnetic field variables.

## V. DISCUSSION

### A. Different limits

In the following, we shall check some known limits using the above exact relation:

(i) When there is no external magnetic field i.e. $\mathbf{H} = 0$, then the exact relation have only term $\langle \delta(\mathbf{u} \times \mathbf{\Omega}) \cdot \delta\mathbf{u} - I\delta\boldsymbol{\omega} \cdot \delta(\mathbf{u} \cdot \nabla\boldsymbol{\omega}) \rangle$, which shows that in absence of magnetic field ferrofluid behave like a simple fluid. If moment of inertia ($I$) is very small then the term $I\delta\boldsymbol{\omega} \cdot \delta(\mathbf{u} \cdot \nabla\boldsymbol{\omega})$ is negligible, so eqn. (38) becomes, $\varepsilon = \frac{1}{2} \langle \delta(\mathbf{u} \times \mathbf{\Omega}) \cdot \delta\mathbf{u} \rangle$, which follow the kolmogorov theory of turbulence.

(ii) Similarly for zero magnetic moment this fluid behaves like simple fluid.

(iii) When magnetic moment of particles aligns towards external magnetic field one can write $\mathbf{M} \propto \mathbf{H}$ or $\mathbf{M} = \alpha\mathbf{H}$, here $\alpha$ is proportionality variable, then $\mathbf{M} \times \mathbf{H} = 0$. If $\alpha$ is constant, then $\langle \delta\mathbf{u} \cdot \delta((\mathbf{M} \cdot \nabla)\mathbf{H}) \rangle = 0$. Hence in eqn. (38) only terms $\delta(\mathbf{u} \times \mathbf{\Omega}) \cdot \delta\mathbf{u}$, $\delta\boldsymbol{\omega} \cdot \delta(\mathbf{u} \cdot \nabla\boldsymbol{\omega})$, $\delta\mathbf{H} \cdot \delta((\mathbf{u} \cdot \nabla)\mathbf{M})$ and $\delta\mathbf{H} \cdot \delta(\boldsymbol{\omega} \times \mathbf{M})$ left in which the last two term represents the settling of magnetic particles.

### B. Other features and perspective

We see that the $\mathbf{M} \times \mathbf{H}$, $\mathbf{M} \times \boldsymbol{\omega}$ and Lamb vector $\mathbf{u} \times \mathbf{\Omega}$ play the key role in energy cascading. For different types of aligned flow states, some or all of them can vanish. For example, for Beltrami flows where the fluid velocity $\mathbf{u}$ and fluid vorticity $\mathbf{\Omega}$ are collinear, the contribution of $\delta(\mathbf{u} \times \mathbf{\Omega})$ in eqn (38) goes away and same will apply for other aligned cases where $\mathbf{M} \parallel \mathbf{H}$ which means that the total magnetization vector is aligned to the external magnetic field and also for $\mathbf{M} \parallel \mathbf{\Omega}$ where the individual particle spin direction is aligned to the total magnetization. It is therefore clear that with each type of alignment, there is a partial suppresion of the turbulent contribution to the energy flux rate. A future work can be proposed to derive this type of exact relation for other type of magnetization equation, using a time varying external magnetic field and also adding compressibility to the ferrofluid. This type of relations can be verified using properly designed laboratory experiments or numerical simulations.

---